\documentclass[preprint,preprintnumbers,amsmath,amssymb,prb]{revtex4}
\usepackage{dcolumn}  
\usepackage{bm}  
\usepackage{natbib}  
\usepackage{amssymb}  
\usepackage{epsfig}  
\usepackage{graphics}  
\usepackage{graphicx}  
\usepackage{color}  
\usepackage{graphicx,epsf,fancyhdr}

\begin{document}

\title{Stability of calcium and magnesium carbonates at 
	lower mantle thermodynamic conditions}

\author{Samuel S. M. Santos$^\dag$, Michel L. Marcondes$^\dag$, 
	Jo\~ao F. Justo$^\ddag$, and Lucy V. C. Assali$^\dag$}  

\affiliation{$\dag$ Instituto de F\'{\i}sica,  
	Universidade de S\~ao Paulo,\\  
	CEP 05508-090, S\~ao Paulo, SP, Brazil \\  
	$\ddag$ Escola Polit\'ecnica, Universidade de S\~ao Paulo,\\  
	CP 61548, CEP 05424-970, S\~ao Paulo, SP, Brazil}  
\date{\today}   

\begin{abstract}
	  
We present a theoretical investigation, based on ab initio calculations and the
quasi-har\-mon\-ic approximation, on the stability properties of magnesium 
(MgCO$_3$) and calcium (CaCO$_3$) carbonates at high temperatures 
and pressures. The results indicate that those carbonates should be stable in 
Earth's lower mantle, instead of dissociating into other minerals, in 
chemical environments with excesses of SiO$_2$, MgO, or MgSiO$_3$. Therefore, 
considering the lower mantle chemical composition, filled with major 
minerals such as MgSiO$_3$ and MgO, calcium and magnesium carbonates are 
the primary candidates as carbon hosts in that region. For the
thermodynamic conditions of the mantle, the results also indicate that 
carbon should be primarily hosted on MgCO$_3$. Finally, the results
indicate that carbon, in the form of free CO$_2$, is unlikely in the lower 
mantle. 
\\
\\
\end{abstract}

\maketitle

\section{Introduction} 
\label{sec1}

Carbon is a unique chemical element, mainly due to its rich bonding nature, 
which provides a wide range of stable and metastable structures in several 
hybridizations and configurations. Particularly, the role of carbon on Earth's 
natural phenomena has been extensively studied over the last few decades. The 
carbon cycle affects atmosphere, oceans and other shallow crustal phenomena, 
directly influencing climates and ecosystems, and consequently life on Earth. 
While there is considerable accumulated knowledge on the carbon cycle near 
Earth's surface, there is still scarce information on the processes associated with 
its deep layers \cite{Hazen2013}.


In order to build consistent models on the Earth's carbon cycle, it is important
to establish a proper understanding on its chemical composition. The knowledge 
on the solar system composition, based on information from carbonaceous 
chondritic meteorites that have hit the Earth, allows to estimate the expected 
amount of carbon that should be present on Earth \cite{Marty2013}. However, the 
current known carbon reservoirs on its shallow layers is about two orders of 
magnitude smaller than the expected value, suggesting that this missing carbon 
should be stored on its deep layers, which has been labelled as deep carbon 
reservoirs \cite{Hazen2013}. Estimates indicate that the Earth's deep interior may 
contain as much as 90\% of all available carbon. Such conclusions 
have been supported by a number of indirect evidences, such as the presence of 
CO$_2$ in magmas \cite{Marty2013} and mantle mineral inclusions in natural 
diamonds \cite{Walter2011,Pearson2014,Maeda2017}.


There are several questions on the properties of deep carbon that remain 
open, such as determining the amount and distribution of those carbon 
reservoirs within major mantle minerals, and understanding the complete carbon 
cycle, associated with exchange of carbon between Earth's surface and its deep 
interior. For example, current estimates suggest that the carbon flux into 
Earth's mantle through subduction substantially exceeds the carbon 
flux emitted by volcanoes, opening the question on the minerals 
that host such carbon in the mantle.


Those issues could only be addressed with a deep 
understanding on the behavior of carbon-related minerals under high pressures 
and temperatures, particularly on the thermodynamic conditions of the 
mantle \cite{Oganov2013}. Although some carbon may be stored in the core, the mantle 
is thought to be its largest reservoir. It is reasonable to assume that a certain 
amount of carbon could be dissolved in silicates, such as MgSiO$_3$, which is by far 
the most abundant mineral in the lower mantle. Therefore, even if this mineral had 
a low to moderate solubility for carbon, it could still be the largest carbon 
reservoir in that region. However, it has been experimentally shown that the 
carbon solubility in silicates is very small, indicating that most 
of the Earth's carbon must be stored in other minerals \cite{Brenker2007}.
Furthermore, silicates are large diffuse reservoirs, which contrasts with the 
concentrated phases of the deep carbon identified as magma carbonates \cite{Jones2013} 
and diamonds \cite{Shirey2013}, as well as carbon species emitted by 
volcanoes \cite{Burton2013}. 


In the lower mantle, it is still not known if carbon exists in a reduced state, 
such as in the form of diamond, or in an oxidized one, such as in the form of 
carbonates. To further explore the potential carriers of carbon, it is 
important to explore the physical properties of those minerals at lower mantle 
thermodynamic conditions, particularly in terms of carbon-related stable phases. 
Several theoretical \cite{Marcondes2016,Oganov2008,Pickard2015} and experimental 
\cite{Fiquet2002,Isshiki2004} investigations have explored the 
high-pressure stability of major carbonates, such as MgCO$_3$, CaCO$_3$, and 
MgCa(CO$_3$)$_2$. However, there is still scarce information on those properties at 
high temperatures \cite{Gavryushkin2017,Smith2018}. Particularly for the lower mantle, 
such conditions would mean pressures up to 140 GPa and temperatures up to 3000 K.


This investigation explores the stability of carbonates, particularly 
MgCO$_3$ and CaCO$_3$, at lower mantle conditions. Here, we did not explore the 
MgCa(CO$_3$)$_2$ mineral, since it is well established that it dissociates into 
MgCO$_3$ and CaCO$_3$ for pressures higher than a few GPa \cite{Shirasaka2002}.
The results were obtained by first principles total energy calculations, combined 
with the quasi-harmonic approximation, in order to obtain the respective Gibbs free 
energies of those minerals in different crystalline phases, at high temperatures 
and pressures. We explored the stability of those carbonates, taking into account a 
number of crystalline phases that have been identified by recent theoretical and 
experimental investigations. Our results indicate that, at high temperatures and 
high pressures, MgCO$_3$ and CaCO$_3$ should be stable against dissociation into other 
minerals at several mantle conditions. Considering the chemical composition 
of the lower mantle, with major concentrations of MgSiO$_3$ and MgO, calcium and 
magnesium carbonates should be 
the primary candidates for carbon hosts in the lower mantle.
However, in the thermodynamic conditions of the mantle, along its geotherm,
magnesium carbonate is more favorable than calcium carbonate.  Additionally, 
the results suggested that carbon in the form of isolated CO$_2$ is unlikely in the 
lower mantle. 

\section{Methods} 
\label{sec2}

\subsection{Ab initio calculations}
\label{sec2a}

The first principles calculations were performed using the Quantum ESPRESSO
computational package \cite{Giannozzi2009}. The electronic interactions were described
within the density functional theory, considering the exchange-correlation (XC)
potential based on the local density approximation (LDA) functional \cite{Kohn1965}.
This functional has been widely used to obtain both static and
dynamic properties of minerals of the Earth's mantle,
although many authors have investigated those properties 
with the generalized gradient approximation (GGA) \cite{Perdew1996}.
It is well established in the literature that static calculations with the LDA 
underestimate the mineral lattice parameters (generally by about 1 to 2 \%) 
and overestimate the respective phase transition pressures and elastic constants 
when compared to experimental values at finite temperatures, while 
calculations with the GGA provide the opposite effects \cite{Michel2018}.
This investigation used only the LDA functional, since by incorporating thermal
effects later on, the lattice parameters increase (and elastic constants decrease), 
going with the appropriate trend toward the respective experimental values.


The electronic wave functions were expanded using the projected augmented wave
(PAW) method \cite{Blochl1994}, with a plane-wave cutoff of 1200 eV.
The valence electronic configurations were described with
($3s^2$ $3p^6$ $4s^2$ $3d^0$ $4p^0$ ) for calcium, ($2s^2$ $2p^2$) for carbon, 
($2s^2$ $2p^6$ $3s^2$ $3p^0$) for magnesium, ($3s^2$ $3p^2$) 
for silicon, and ($2s^2$ $2p^4$) for oxygen \cite{Holzwarth2001}. The Brillouin zones 
for electronic states of crystalline phases were sampled by a 
$4 \times 4 \times 4$ k-mesh for the carbonates and silicates, 
a $8 \times 8 \times 8$ k-mesh for alkaline earth oxides (MgO and CaO), and 
a $6 \times 6 \times 6$ k-mesh for SiO$_2$ and CO$_2$, in order to provide 
an approximately equivalent density of k-points for all materials. 

Strict convergence criteria were taken into account for the simulations, with
the atomic positions being considered converged when all forces acting
on atoms were smaller than 0.01 eV \AA$^{-1}$.
For each pressure, from 0 to 150 GPa, the structures were optimized
using the damped variable cell shape molecular dynamics method \cite{Wentzcovitch1993}. 
Then, a second order Birch-Murnaghan equation of state was used to fit the 
compression data.

\subsection{Thermodynamic properties}
\label{sec2b}

The thermodynamic properties were investigated by computing the vibrational 
modes (phonons) of the crystals using the Density Functional Perturbation 
Theory \cite{Baroni2001}, with a $12 \times 12 \times 12$ q-mesh to calculate the 
vibrational density of states (VDOS) \cite{Wang2010}, and the quasi-harmonic 
approximation (QHA) \cite{Wentzcovitch2010} to compute the respective Gibbs free energies.

The results for finite temperatures were reported within the validity range of 
the QHA \cite{Wentzcovitch2010}. It is well established in the literature that the QHA results for the thermal expansion, $\alpha(\text{P,T})$, start diverging beyond a certain temperature, in opposition to available experimental results. Such divergence results from the fact that this methodology disregards anharmonic effects \cite{Wentzcovitch2010,Wentzcovitch2004}. However, this quantity is  also clearly quite sensitive to the choice of XC functional used  \cite{Michel2018}, and could still be used at high temperatures, providing results consistent with experimental data in the region in which the thermal expansion coefficient does not diverge.
Therefore, at high temperatures, the validity region of the QHA has been 
established  \cite{Wentzcovitch2010,Carrier2007} as
$[{\partial^2 \alpha(\text{P,T})}/{\partial \text{T}^2}]_{\text{P}} \leq 0$.

\begin{figure}[!h]
\centering
\includegraphics[scale=0.4, angle=270]{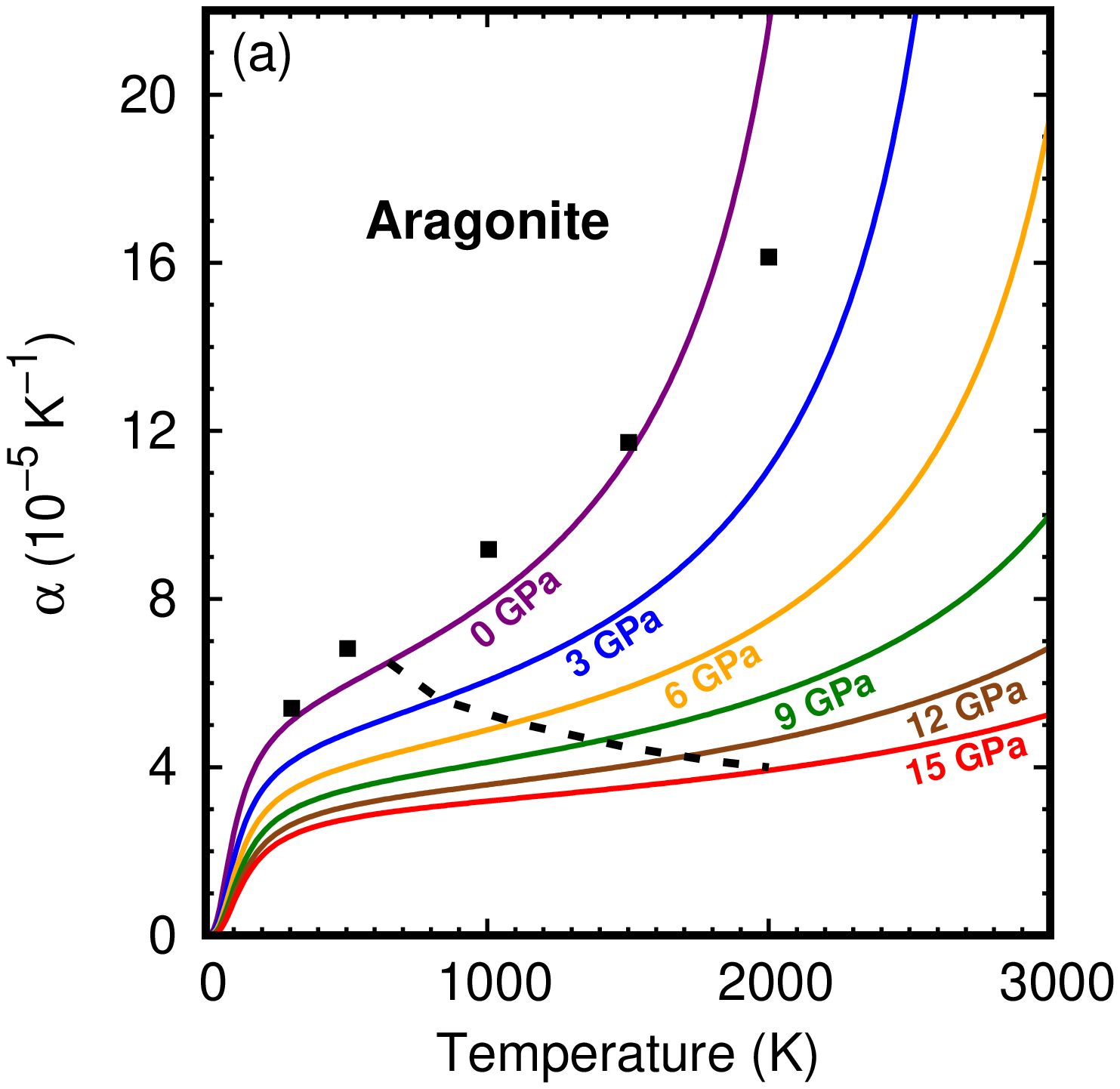}
\includegraphics[scale=0.4, angle=270]{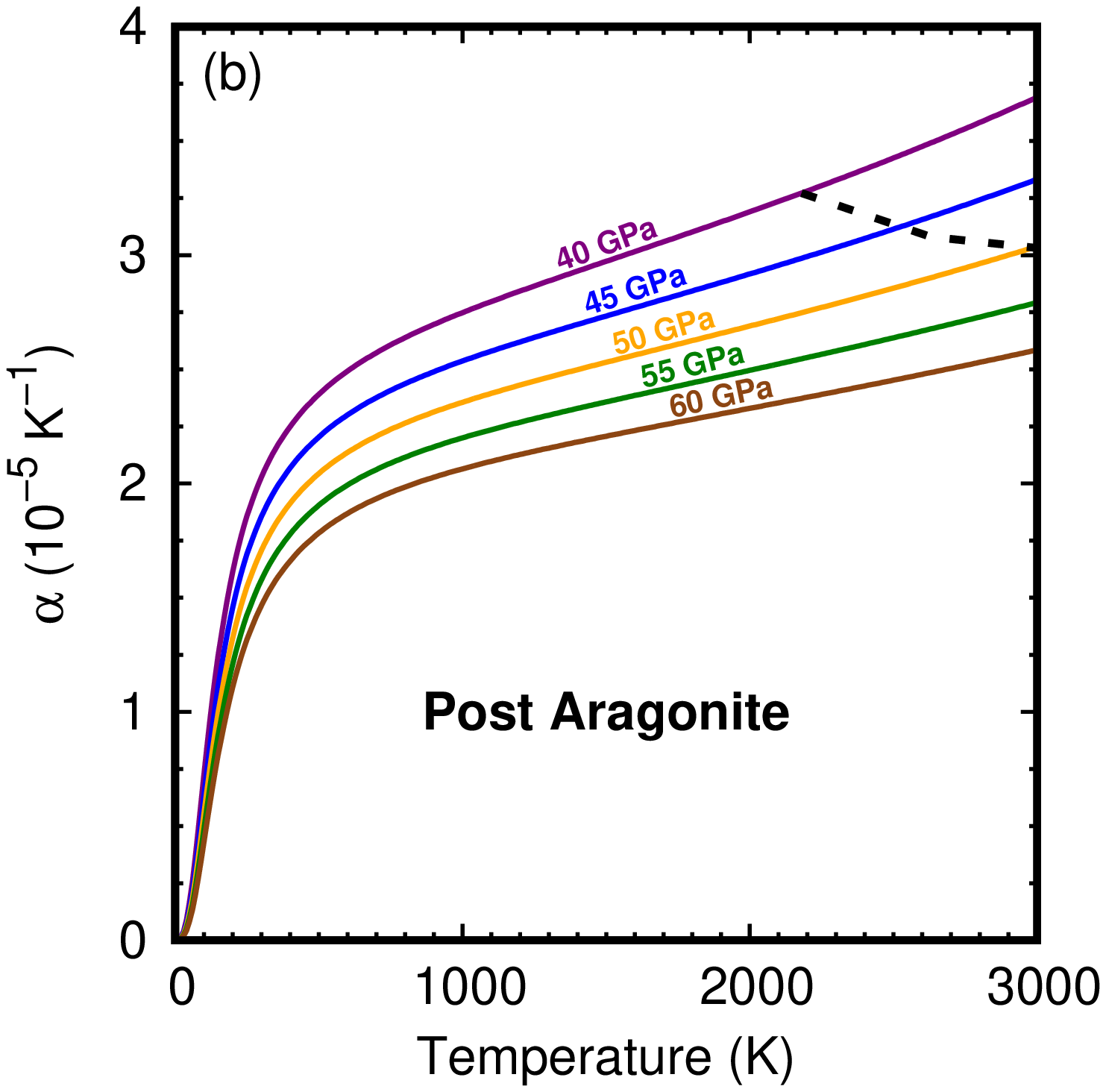}
\caption{Thermal expansion coefficient at several pressures as a function 
of temperature for CaCO$_3$ in (a) aragonite and (b) post-aragonite phases. 
The QHA boundary is defined by the position of the inflection points of 
$\alpha(\text{P,T})$, discussed in section \ref{sec2b}. 
Experimental results at 0 GPa are presented with black symbols \cite{Litasov2017}.}
\label{fig1}
\end{figure}

Figure \ref{fig1} shows the thermal expansion coefficient at several pressures
for CaCO$_3$ in aragonite and post-aragonite phases, which represents a stringent
test for this methodology. The figure shows the respective validity regions of the 
QHA for those phases, indicating that at very high pressures, the QHA 
is still valid at temperatures up to 3000 K. The figure also shows that
within the validity limit of the QHA, the  thermal expansion coefficient
is in good agreement with available experimental data \cite{Litasov2017}. 

We explored the validity of the QHA at the lowest pressure phases for 
all minerals considered in this investigation. The region
of low pressures and high temperatures represents a stringent test for 
the QHA methodology, since  it is when the thermal expansion coefficients 
present major divergences when compared to experimental data.
The QHA provided appropriate thermal expansion coefficients under
the thermodynamic conditions of interest in this investigation.

Our results on thermal properties of all minerals investigated here 
are within the validity region of the QHA, and in good agreement with 
available theoretical and experimental data 
\cite{Litasov2017,Wentzcovitch2010,Dorogokupets2007,Li2006,Karki2000}.


\subsection{Crystalline phases}
\label{sec2c}

In order to explore the stability of carbon-related minerals, within the 
thermodynamic conditions of the lower mantle, we initially studied several
crystalline phases of CaCO$_3$ and MgCO$_3$ carbonates and CaSiO$_3$ and MgSiO$_3$ 
silicates in a wide pressure range, by computing their respective enthalpies and 
Gibbs free energies. 

For any mineral at a certain temperature and pressure, the respective stable phase 
is determined as the one with the lowest Gibbs free energy value, using the 
theoretical model described in the previous section. 
Additionally, a mineral follows a phase transition at a certain pressure when the free energy of the stable phase becomes higher than the one of a different phase. Our results on stability and phase transition pressures of several carbonate and silicate minerals were in good 
agreement with results from other theoretical and experimental investigations 
\cite{Marcondes2016,Oganov2008,Pickard2015,Isshiki2004}.


We considered a number of crystalline phases for CaCO$_3$, MgCO$_3$, MgSiO$_3$, 
CaSiO$_3$, MgO, CaO, SiO$_2$, and CO$_2$. First of all, MgO was considered only in 
the Fm$\overline{3}$m phase, since another investigation has shown that it remains 
in that phase up to 227 GPa \cite{Duffy1995}. CO$_2$ was 
considered only in the I$\overline{4}$2d phase \cite{Datchi2012}.
We performed static calculations for several other CO$_2$ phases and we obtained
that the above mentioned phase is the most stable for pressures above 10 GPa,
which is fully consistent with results of another investigation that found this phase
for pressures from 19 to 150 GPa \cite{Oganov2008}. Additionally, ignoring those
other low pressure stable phases does not compromise our conclusions on the 
properties of Earth's lower mantle.

For CaSiO$_3$, we considered only the tetragonal (I/4mnm) phase for all the 
pressures and temperatures studied here \cite{Shim2002}, since it is the 
most important phase in the thermodynamic conditions of interest.
Moreover, within such conditions, our methodology provides results which comply with
the validity criteria described in section \ref{sec2b}, i.e.   
the thermal expansion coefficient of this phase does not diverge up to 3000 K.

\begin{figure}[!h]
\centering
\includegraphics[scale=1]{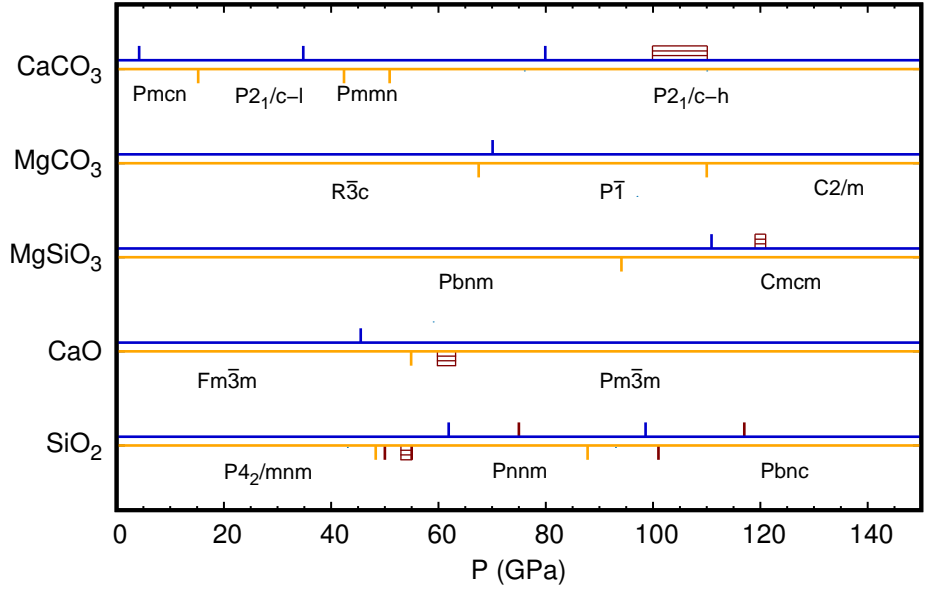}
\caption{Stable crystalline phases, as 
function of pressure, of CaCO$_3$, MgCO$_3$, MgSiO$_3$, 
CaO, and SiO$_2$ materials considered in this investigation,
computed with theoretical approximations presented in section
\ref{sec2a}. The  figure presents results for two temperatures: 
300 K (orange line) and 2000 K (blue line), with the respective 
transitions represented by vertical lines. Experimental values 
for phase transitions, with respective experimental error bars, 
are presented with brown symbols for: CaCO$_3$ \cite{Lobanov2017}, 
MgSiO$_3$ \cite{Murakami2004},  CaO \cite{Yamanaka2002}, 
and SiO$_2$ \cite{Hemley2000,Andrault1998,Ono2002,Murakami2003}.}
\label{fig2}
\end{figure}

Figure \ref{fig2} shows the stable crystalline phases for all other minerals 
considered in this investigation, as function of pressure at temperatures of 300 
and 2000 K, in which there are available experimental data for comparison.
For CaCO$_3$ at 300 K between 0 and 150 GPa, the material goes from Pmcn to 
P2$_1/$c-l at 15 GPa, from P2$_1/$c-l to Pmmn at 42 GPa, and from Pmmn to 
P2$_1/$c-h at 51 GPa, results which are consistent with 
another theoretical investigation \cite{Pickard2015}. Figure \ref{fig2} also shows 
that, as the temperature increases, there is a major change on the transition 
pressures. At a high temperature, 2000 K, the  Pmmn to P2$_1/$c-h phase transition 
occurs at about 80 GPa, in agreement with the theoretical and experimental
values of 76 GPa \cite{Pickard2015} and $105\pm 5$ GPa \cite{Lobanov2017}, respectively.

For MgCO$_3$ at 300 K, the material goes from R$\overline{3}$c to P$\overline{1}$
at 68 GPa, and from P$\overline{1}$ to C2$/$m at 110 GPa, all results in good 
agreement with other theoretical investigations using static calculations \cite{Pickard2015}. 
Our results indicate that at temperatures over 1850 K, this mineral follows a direct 
transition from R$\overline{3}$c to C2$/$m, i.e. above that temperature 
the P$\overline{1}$ is not stable at any pressure. 
Therefore, our results indicate that P$\overline{1}$ should be of 
low geophysical interest for studies of the lower mantle properties.

For MgSiO$_3$ at 300 K, the mineral goes from Pbnm to Cmcm at 94 GPa, which is 
consistent with another theoretical investigation \cite{Tsuchiya2004}. At 2000 K,
this transition occurs at 111 GPa, in good agreement with the experimental 
value of $120\pm 3$ GPa \cite{Murakami2004}.

For CaO at 300 K, the material goes from Fm$\overline{3}$m to Pm$\overline{3}$m.
The transition between these two phases is found at 55 GPa, in good agreement with 
experimental data that identified this transition between 59.8 and 
63.2 GPa \cite{Yamanaka2002}.

For SiO$_2$ at 300 K (2000 K), the material goes from P4$_2$/mnm to Pnnm at 48 GPa 
(62 GPa), and from Pnnm to Pbnc at 88 GPa (99 GPa). Those transition values are in 
good agreement with experimental data at 
300 K \cite{Hemley2000,Andrault1998} and 2000 K \cite{Ono2002,Murakami2003}.

\section{Results}
\label{sec3}

\subsection{Decomposition of carbonates at high pressures and temperatures}
\label{sec3a}

Most of the Earth's oxidized carbon is expected to be harbored by Mg and/or Ca 
carbonate forms under mantle pressures and temperatures. In order to identify 
the potential carbon hosts in this region, we initially explore the energetics
associated with the following decomposition reactions:
\begin{eqnarray*}
\hspace{3cm} \rm CaCO_3 & \Rightarrow &  \rm CaO + CO_2 \hspace{3cm}  (R1) \\
\hspace{3cm} \rm MgCO_3 & \Rightarrow & \rm MgO + CO_2  \hspace{3cm} (R2)
\end{eqnarray*}

\begin{figure}[!h]
\centering
\includegraphics[scale=1]{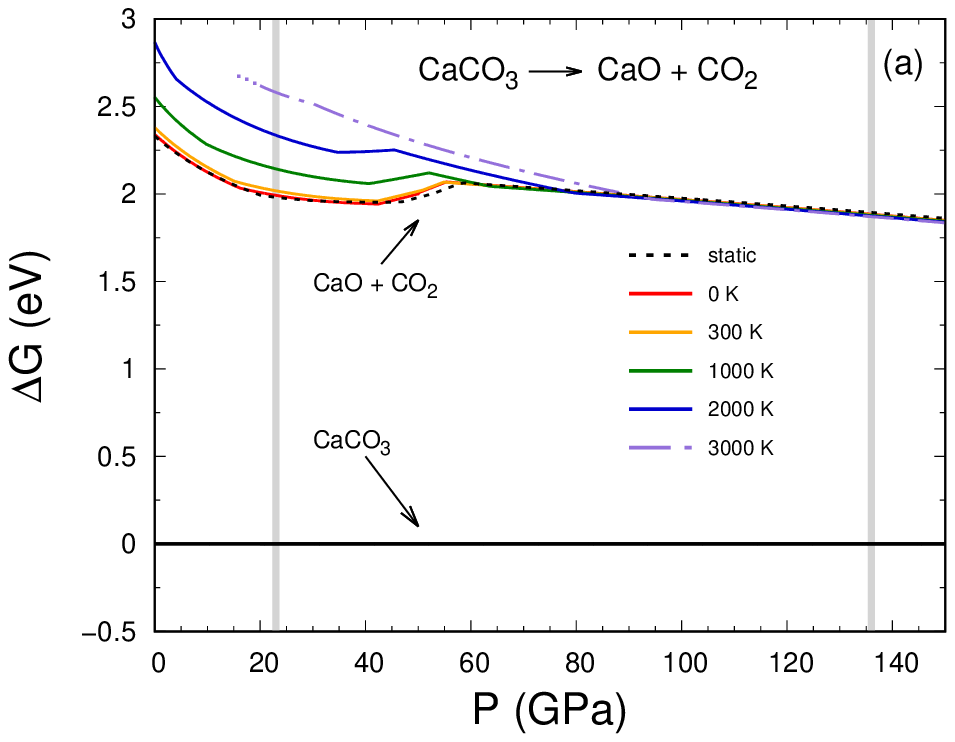}
\includegraphics[scale=1]{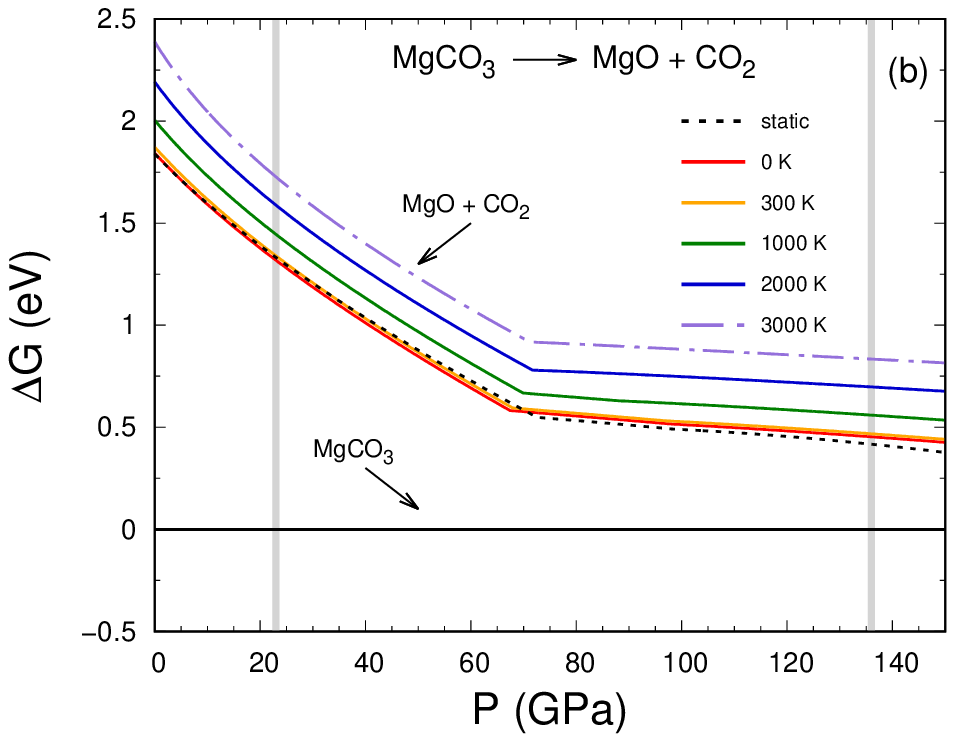}
\caption{The relative Gibbs free energies per u.f. as function of 
pressure, at several temperatures, for (a) CaCO$_3$ $\rightarrow$ CaO + CO$_2$ and (b) 
MgCO$_3$ $\rightarrow$ MgO + CO$_2$ reactins. The dotted black lines represent the 
respective relative enthalpies. The vertical gray lines indicate 
the pressures at the top (23 GPa) and bottom (136 GPa) of the lower mantle. 
The kinks in the curves arise from phase transitions that occur in the minerals 
described in section \ref{sec2c} and shown in figure \ref{fig2}, 
which are used to explore the dissociation reactions.}
\label{fig3}
\end{figure}

Figure \ref{fig3} shows the relative Gibbs free energy per unit formula (u.f.) 
as a functions of pressure and temperature. It shows that the direct decompositions of 
CaCO$_3$ and MgCO$_3$ into their respective alkaline earth oxides plus CO$_2$ are 
unfavorable all over the lower mantle. CaCO$_3$ and MgCO$_3$ 
show similar trends with pressure, i.e., increasing the pressure reduces the 
Gibbs free energy difference for decomposition, which are positive
for all the pressures of interest of the lower mantle. However, the energy cost for 
the CaCO$_3$ reaction is much higher than the one for MgCO$_3$.

It should be stressed that our static results, black lines in 
figure \ref{fig3}, are in very good agreement with the static results 
presented in other recent investigations \cite{Oganov2008,Pickard2015}, as well as with 
experimental data \cite{Fiquet2002}. Those investigations, without taking into 
account temperature effects, have suggested that free CO$_2$ does not occur as an 
independent phase within the Earth's mantle. According to the results presented in
figures \ref{fig3}(a) and (b), temperature effects do not alter the relative stability 
of carbonates, when compared to their most elementary constituents. In reality, the 
phenomenology is quite the contrary, as the figure shows, a temperature increase 
further reduces the possibility of free CO$_2$ to exist in that region.

\subsection{Stability of carbonates under excess of \textnormal{SiO}$_2$}
\label{sec3b}

We now evaluate the stability of carbonates in a condition of excess of SiO$_2$, as 
represented by reactions (R3) and (R4). This condition is particularly important when 
one takes into account that the upper and lower mantle have material mixing, resulting 
from the basaltic part of subducting slabs, which is rich in SiO$_2$ \cite{Oganov2008}. 

\begin{eqnarray*}
\hspace{3cm} \rm MgCO_3 + SiO_2 & \Rightarrow & \rm MgSiO_3 + CO_2  \hspace{3cm}  (R3) \\
\hspace{3cm} \rm CaCO_3 + SiO_2 & \Rightarrow & \rm CaSiO_3 + CO_2  \hspace{3cm} (R4)
\end{eqnarray*}

\begin{figure}[!h]
\centering
\includegraphics[scale=1]{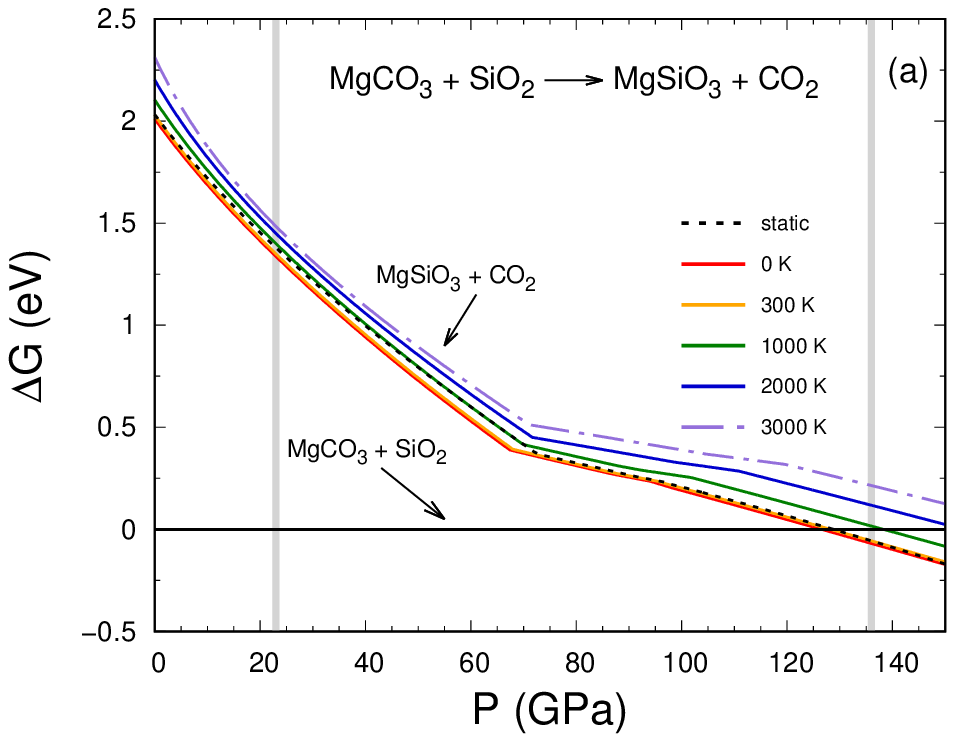}
\includegraphics[scale=1]{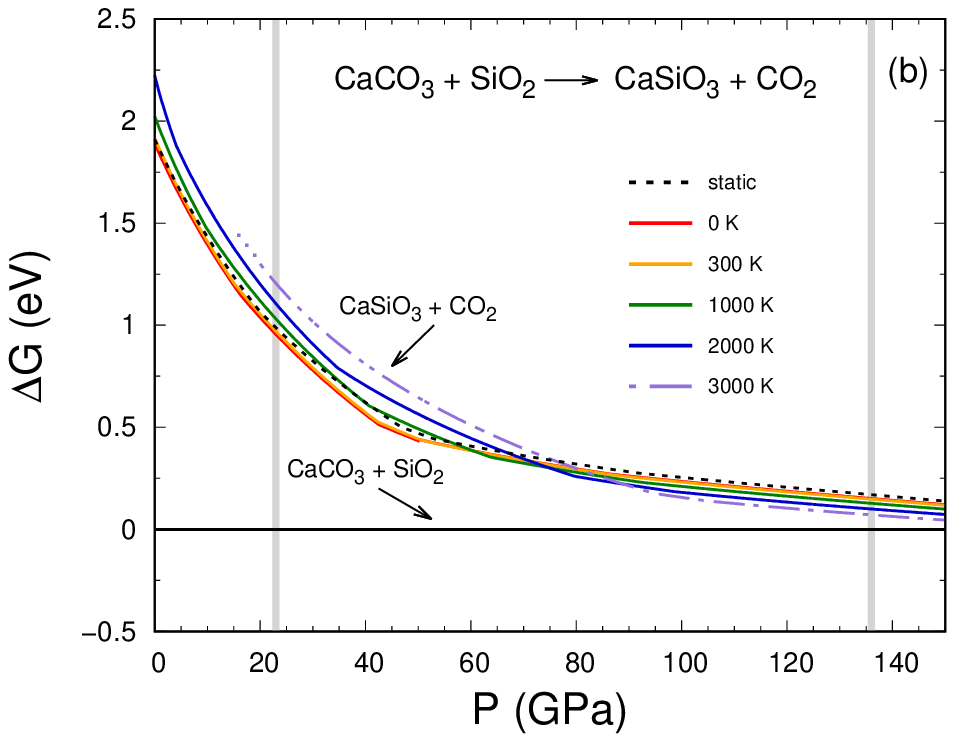}
\caption{The relative Gibbs free energy per u.f. as function of pressure,
at several temperatures, for (a) MgCO$_3$ + SiO$_2$ $\rightarrow$ MgSiO$_3$ + CO$_2$  
and (b) CaCO$_3$ + SiO$_2$ $\rightarrow$ CaSiO$_3$ + CO$_2$  reactions. 
The dotted lines represent the relative enthalpies for the reactions. 
The vertical gray lines indicate the 
pressures at the top and bottom of the lower mantle.}
\label{fig4}
\end{figure}

Figure \ref{fig4} shows the relative Gibbs free energies for Mg and Ca carbonates to 
transform into their respective silicates. The results indicated that reactions 
(R3) and (R4) are mostly unfavorable, carbonates do not react to form silicates within 
those thermodynamic conditions. Therefore, there should be no CO$_2$ formation in the 
lower mantle as result of those reactions. Static results, in 
figure \ref{fig4}(a) indicate that reaction (R3) is unfavorable up to 127 GPa,
i.e. MgCO$_3$ is more stable than MgSiO$_3$. This pressure is lower than the one in the
lower mantle-core boundary of 136 GPa, such that static results suggested that
reaction (R3) would be favorable at the bottom of the lower mantle, which could lead 
to the generation of free CO$_2$. The figure shows that reaction
(R3) becomes less favorable with increasing temperature, such that
this reaction seems unfavorable at typical temperatures of the lower mantle bottom.

Our static results for reaction (R3), black line in 
figure \ref{fig4}(a), are in good agreement with static results 
presented in  recent theoretical investigations \cite{Oganov2008,Pickard2015}.
However, the transition pressure found here is lower than the one 
found by those authors. Such differences could be explained by the functionals used 
to describe electron-electron interactions, since our investigation used the LDA 
\cite{Kohn1965}, while those investigations use the GGA  \cite{Perdew1996}. The choice of 
LDA over GGA in this investigation was discussed in detail in section \ref{sec2a}.

According to figure \ref{fig4} (b), the reaction (R4) indicates that CaCO$_3$ is more 
stable than CaSiO$_3$ up to 150 GPa at any temperature. Since this pressure is much 
larger than the ones at the bottom of the lower mantle, this reaction would not occur 
in that region. Our static results on the transition pressure for reaction (R4) 
disagree with another theoretical investigation \cite{Oganov2008}, which predicted a
reaction of calcium carbonate with SiO$_2$ within the pressure range of the lower 
mantle. In fact, that investigation did not take into account the P2$_1/$c-h
phase of CaCO$_3$, which may explain the differences in the conclusions.
On the other hand, our static results are in good agreement with the ones 
of a more recent investigation \cite{Pickard2015}.

It should be pointed out that our investigation 
carries some uncertainties on the transition pressures of SiO$_2$, as shown
in figure \ref{fig2}, which could play some role on the final conclusions about 
the reactions (R3) and (R4), particularly on the properties near
the core-mantle boundary.

\subsection{Stability of carbonates under excess of \textnormal{MgO} or \textnormal{MgSiO}$_3$}
\label{sec3c}

It is reasonably well established that MgSiO$_3$ and MgO are the two major minerals 
in the lower mantle. Therefore, it is important to study the stability of MgCO$_3$ 
or CaCO$_3$ carbonates in a rich environment with those two major minerals. 
To explore those conditions, we evaluate the energetics associated to reactions 
(R5) and (R6).

\begin{eqnarray*}
\hspace{3cm}  \rm MgCO_3 + CaO  & \Rightarrow & \rm  CaCO_3 + MgO  \hspace{3cm} (R5) \\
\hspace{3cm} \rm CaCO_3 + MgSiO_3  & \Rightarrow & \rm MgCO_3 + CaSiO_3 \hspace{3cm} (R6) 
\end{eqnarray*}

\begin{figure}[!h]
\centering
\includegraphics[scale=1]{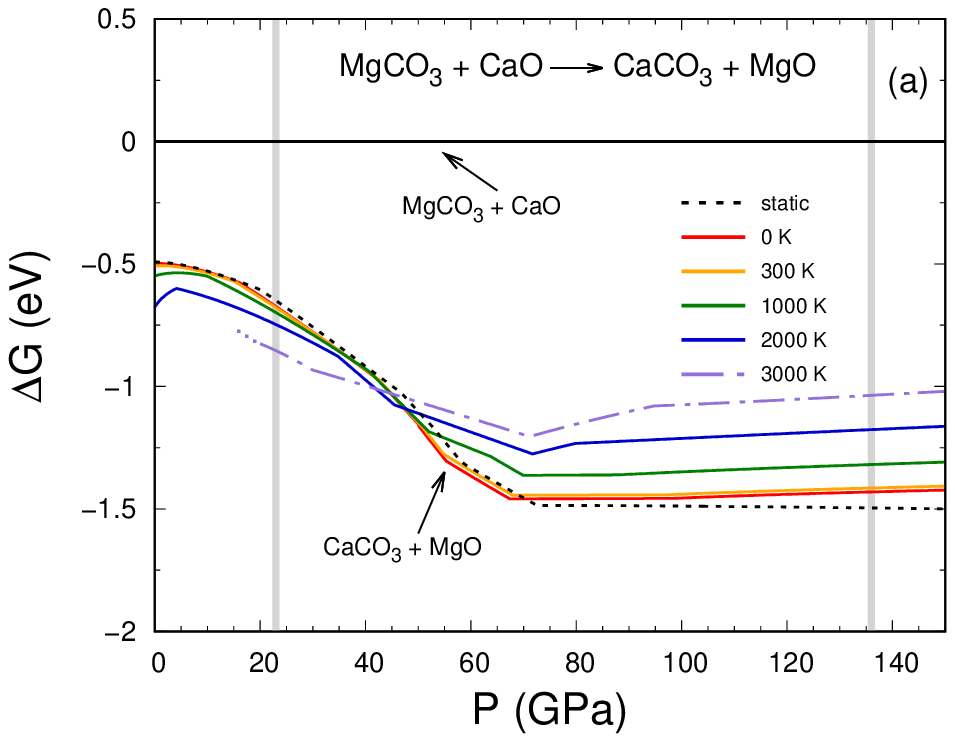}
\includegraphics[scale=1]{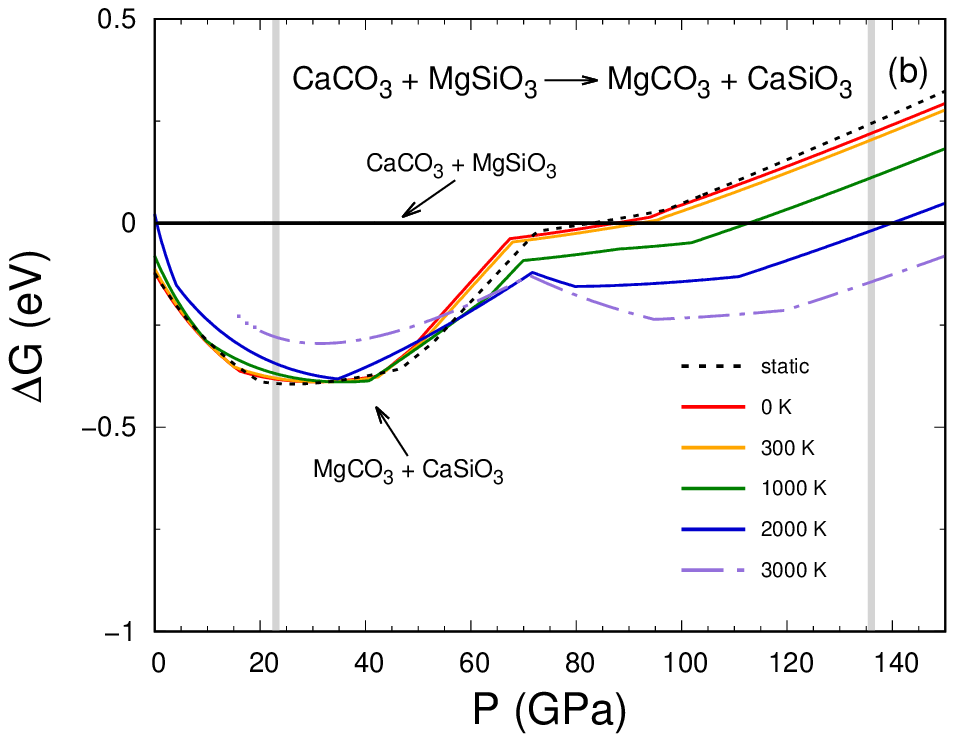}
\caption{The relative Gibbs free energy per u.f. as function of pressure,
at several temperatures, for (a) MgCO$_3$ + CaO  $\rightarrow$ CaCO$_3$ + MgO and 
(b) CaCO$_3$ + MgSiO$_3$ $\rightarrow$ MgCO$_3$ + CaSiO$_3$ reactions. 
The black dotted lines represent 
the relative enthalpies in (a) and (b). The vertical gray lines indicate 
the pressures at the top and bottom of the lower mantle.}
\label{fig5}
\end{figure}

Figure \ref{fig5}(a) shows that the reaction (R5) is energetically favorable, i.e. 
CaCO$_3$ + MgO is more stable than MgCO$_3$ + CaO, at lower mantle conditions. Although 
by increasing the temperature, the relative Gibbs free energy is reduced, this effect 
is not large enough to change the stability of MgCO$_3$. Therefore, when there is 
MgO in excess, as expected in a pyrolitic mantle, CaCO$_3$ is the stable carbonate under 
those thermodynamic conditions. Additionally, our static results are in good 
agreement with recent results from other investigations \cite{Oganov2008,Pickard2015}.

We now explore the conditions of MgSiO$_3$ excess, which is the main mineral in the 
lower mantle. Figure \ref{fig5}(b) shows that reaction (R6) presents a much richer 
phenomenology than reaction (R5) across the pressure range of interest. Our static 
calculations show that  MgCO$_3$ + CaSiO$_3$ is more favorable than
CaCO$_3$ + MgSiO$_3$ at low pressures, but at pressures higher than 75 GPa, then 
CaCO$_3$ + MgSiO$_3$ becomes favorable. However, as the figure shows, there is a dramatic 
change in the behavior at high temperatures. An increase in temperature increases 
the pressure in which such reaction could be favorable.

For typical lower mantle temperatures, in the order of at least 2000 K, the reaction 
(R6) is not favorable anymore. Therefore, the results in figure \ref{fig5}(b) indicated 
that, at lower mantle conditions under excess of MgSiO$_3$, carbonates appear to be 
favorable in the form of MgCO$_3$. This conclusion is consistent with 
an experimental investigation \cite{Isshiki2004}, which suggested MgCO$_3$ as the main 
oxidized carbon host in Earth's mantle.

\begin{figure}[!h]
\centering
\includegraphics[scale=1]{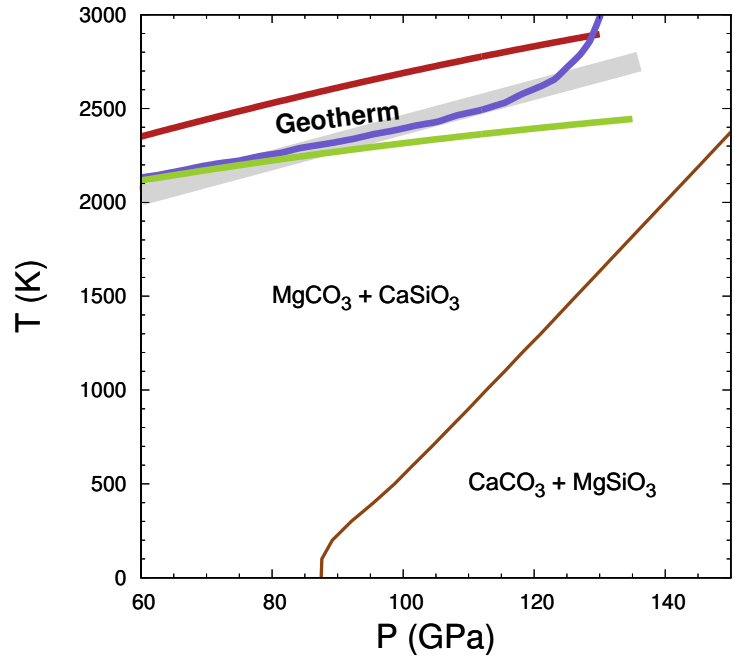}
\caption{Phase diagram for the stability (brown line) 
of MgCO$_3$ + CaSiO$_3$ versus CaCO$_3$ + MgSiO$_3$. 
The figure also shows several geotherms:
thick gray \cite{Isshiki2004}, green \cite{Brown1981}, 
blue \cite{Boehler2000}, and red \cite{Anderson1982} lines.}
\label{fig6}
\end{figure}

Figure \ref{fig6} shows the phase diagram for reaction (R6), along with some Earth's 
geotherms \cite{Isshiki2004,Brown1981,Boehler2000,Anderson1982}. According to the figure, all 
geotherms lie in the region of stability of MgCO$_3$, indicating this mineral as 
the most likely carbon host in the lower mantle, as expected in a pyrolitic mantle. 
However, it should be pointed that there is still controversy on the temperatures at 
the lower mantle bottom, therefore, the stability of CaCO$_3$ over MgCO$_3$ would require a much colder mantle.

\section{Summary} 
\label{sec4}

In summary, this investigation explored the stability of MgCO$_3$ and CaCO$_3$ carbonates in
the thermodynamic conditions of Earth's lower mantle. The results indicated that a direct decomposition of those carbonates is unfavorable at high temperatures and pressures. Assuming an iron free pyrolitic lower-mantle composition, the investigation also explored the stability of those carbonates in conditions of excess of SiO$_2$, MgO, and MgSiO$_3$. 
	
In a MgO-rich environment, the results showed that calcium carbonate is more stable than magnesium carbonate, and the relative energy reduction with increasing temperature was not large enough to change the stability of CaCO$_3$. On the other hand, in a MgSiO$_3$-rich environment, the static results showed that at the upper half of the lower mantle, the MgCO$_3$ + CaSiO$_3$ reaction is more favorable than the CaCO$_3$ + MgSiO$_3$ one, while at pressures higher than 75 GPa, the latter becomes favorable. However, when the effects of temperature
are taken into account, this behavior changes dramatically and the increase in temperature increases the pressure at which this reaction becomes favorable, and finally the magnesium carbonate turns out to be more stable than the calcium carbonate.
	
Since magnesium silicate is the main component of a pyrolitic mantle, it can be inferred that carbonates appear to be favorable in the form of MgCO$_3$. Therefore, the magnesium carbonate should be the main host of oxidized carbon in most of the lower mantle. Only in the bottom of the mantle or in a  region with MgO in excess that calcium carbonate could become preferable. However, in the bottom of the mantle, this carbonate would be favorable only considering geotherms with very low increase in temperature close to the core-mantle boundary.
	
The results also showed that both carbonates do not decompose into their respective alkaline oxides plus CO$_2$ through the entire lower mantle, which indicates low concentration of free carbon dioxide in those regions. However, the decomposition reaction of MgCO$_3$ into MgO+CO$_2$ would only be possible very close to the core-mantle boundary. Furthermore, free CO$_2$ could be produced in an environment with excess of SiO$_2$, as in small silica-rich basaltic parts of the subducted slabs.

All those results add new evidences for the presence of carbon on deep mantle in the form of
carbonates \cite{Hazen2013,Marcondes2016}. However, it is still uncertain how the presence of iron, and its rich phenomenology associated to spin transition at high pressures
\cite{Wentzcovitch2009,Liu2015}, would affect the stability of carbonates in the
lower mantle. Moreover, the decomposition of CO$_2$ into diamond plus oxygen should be explored to enrich the discussion of the presence of carbon in the deep mantle in a reduced state. 


\vspace{0.5cm}
\noindent  
\textbf{Acknowledgments}  
  
This investigation was supported by the Brazilian agencies CNPq and CAPES. We 
acknowledge resources from the Blue Gene/Q supercomputer 
supported by the Center for Research Computing (Rice University) and 
the Superintend\^encia de Tecnologia da Informa\c{c}\~ao 
(Universidade de S\~ao Paulo).
\vspace{0.5cm}

\bibliography{ref}  

\vfill  
\eject


\vfill  
\eject

\vfill  
\eject

\vfill  
\eject  

\vfill  
\eject  

\vfill  
\eject  

\end{document}